\documentclass[aps,prb,twocolumn,floatfix,showpacs,superscriptaddress,fleqn]{revtex4-1}
\usepackage{amssymb}

\usepackage{graphicx}
\usepackage{amsmath}
\usepackage{bm}
\usepackage{color}
\usepackage{braket}

\usepackage[breaklinks]{hyperref}
\usepackage[all]{hypcap}

\hypersetup{
    plainpages=false,
    bookmarks=false,         
    unicode=false,          
    pdfborder={0 0 0}
    pdftoolbar=true,        
    pdfmenubar=true,        
    pdffitwindow=false,     
    pdfstartview={FitH},    
    pdftitle={PRB},    
    pdfauthor={Paulina Plochocka},     
    pdfproducer={LNCMI-CNRS-UGA-UPS-INSA}, 
    pdfkeywords={High} {Magnetic} {Fields}, 
    pdfnewwindow=true,      
    linktoc=section,
    colorlinks=true,       
    linkcolor=blue,          
    citecolor=red,        
    filecolor=magenta,      
    urlcolor=blue           
}

\begin{document}

\author{A.\ A.\ Mitioglu}
\affiliation{Laboratoire National des Champs Magn\'etiques Intenses, CNRS-UGA-UPS-INSA, Grenoble and Toulouse,
France}\affiliation{Institute of Applied Physics, Academiei Str.\ 5, Chisinau, MD-2028, Republic of Moldova}

\author{K. Galkowski}
\affiliation{Laboratoire National des Champs Magn\'etiques Intenses, CNRS-UGA-UPS-INSA, Grenoble and Toulouse, France}

\author{A. Surrente}
\affiliation{Laboratoire National des Champs Magn\'etiques Intenses, CNRS-UGA-UPS-INSA, Grenoble and Toulouse, France}

\author{L. Klopotowski}
\affiliation{Institute of Physics, Polish Academy of Sciences, al. Lotnik\'ow 32/46, 02-668 Warsaw, Poland}

\author{D. Dumcenco}
\affiliation{Electrical Engineering Institute and Interdisciplinary Center for Electron Microscopy (CIME), \'Ecole
Polytechnique F\'ed\'erale de Lausanne (EPFL), CH-1015 Lausanne, Switzerland}

\author{A. Kis}
\affiliation{Electrical Engineering Institute and Interdisciplinary Center for Electron Microscopy (CIME), \'Ecole
Polytechnique F\'ed\'erale de Lausanne (EPFL), CH-1015 Lausanne, Switzerland}

\author{D. K. Maude}
\affiliation{Laboratoire National des Champs Magn\'etiques Intenses, CNRS-UGA-UPS-INSA, Grenoble and Toulouse, France}

\author{P.\ Plochocka}\email{paulina.plochocka@lncmi.cnrs.fr}
\affiliation{Laboratoire National des Champs Magn\'etiques Intenses, CNRS-UGA-UPS-INSA, Grenoble and Toulouse, France}

\title{Magneto-excitons in large area CVD grown monolayer MoS$_{2}$ and MoSe$_{2}$ on sapphire}


\date{\today}

\begin{abstract}
Magneto transmission spectroscopy was employed to study the valley Zeeman effect in large-area monolayer MoS$_{2}$ and
MoSe$_{2}$. The extracted values of the valley g-factors for both A- and B-exciton were found be similar with $g_v
\simeq -4.5$. The samples are expected to be strained due to the CVD growth on sapphire at high temperature
($700^\circ$C). However, the estimated strain, which is maximum at low temperature, is only $\simeq 0.2\%$. Theoretical
considerations suggest that the strain is too small to significantly influence the electronic properties. This is
confirmed by the measured value of valley g-factor, and the measured temperature dependence of the band gap, which are
almost identical for CVD and mechanically exfoliated MoS$_2$.
\end{abstract}

\maketitle



\section{Introduction}

Monolayer transition metal dichalcogenides (TMDs) have recently emerged as an exciting material system in which coupled
spin-valley physics can be
explored.\cite{Wang2012,mak10,spl10,Xiao2012,cao12,mak12,zen12,sal12,jon13,wu13,Xu2014,mak14} Unlike their bulk form,
monolayer TMDs exhibit a direct band gap in a visible spectral range located at two inequivalent $\pm K$
valleys.\cite{mak10,spl10} The spin states are split by a strong spin-orbit interaction, and the order of the spin
states is reversed in the $\pm K$ valleys as a result of time reversal symmetry. Due to the large spin-orbit splitting,
the inter band optical absorption gives rise to well separated A and B-exciton transitions. The selection rules for
these transitions are governed by the orbital magnetic moment resulting from the Bloch part of the carrier wave
function.\cite{yao08} Since the crystal structure of a monolayer TMD lacks an inversion center, the out-of-plane
element of the orbital magnetic moment is nonzero and its sign depends on the valley index. This results in optical
transitions in $\sigma^{\pm}$ polarizations, which involve carriers in the $\pm K$ valleys, providing an access to the
valley index via optical spectroscopy.\cite{cao12,mak12,zen12,sal12} Photoluminescence revealed a large degree of
circular polarization, \cite{cao12,mak12,zen12,sal12,jon13} reaching 100\% for a resonant excitation,\cite{mak12} which
is extremely promising with a view to employing the valley pseudospin degree of freedom in novel applications in,
\emph{e.g.}, quantum information processing.\cite{Xiao2012,cao12,mak12,Zhu2011} In this respect, the development of
large area monolayer TMDs suitable for large scale device applications is crucial.

The existence of a valley-contrasting magnetic moment opens a possibility of controlling the valley pseudospin with an
external magnetic field.\cite{Rose2013,cai13,kor14,ho14,chu14,Mitioglu2015} The application of a magnetic field,
perpendicular to the layer lifts the valley degeneracy splitting the exciton transitions. In monolayer TMDs the
magnetic moment of the carriers has three possible contributions, (i) intracellular $\mu_k = \pm 2\mu_B$ magnetic
moment originating from the orbital contribution of the valence band d-orbitals\cite{liu13}, (ii) the intercellular
valley magnetic moment, which is associated with the Berry curvature,\cite{Xiao2012} and (iii) the spin Zeeman magnetic
moment. As the optical transitions conserve spin, the spin magnetic moment does not contribute to the valley splitting.
In a simple two band model, the masses of the valence and the conduction band are identical so that the intercellular
valley magnetic moment is the same for the valence and conduction bands. Thus, there is no intercellular contribution
to the valley splitting which arises solely from the $\mu_k = \pm 2\mu_B$ angular momentum of the valence d-orbitals
giving a valley g-factor $g_v = -4$, close to the reported values from photoluminescence (PL) studies in transition
metal diselenides.\cite{MacNeill2015,Li2014,Srivastava2015,Aivazian2015,Mitioglu2015,Wang15}.

Surprisingly, a significant deviation from $g_v =-4$ was reported by Aivazian et al.\cite{Aivazian2015}. A systematic
study showed that the valley g factor can take values of either $\simeq -2.8$ or $\simeq -1.6$, depending on the
sample. This was attributed to asymmetry between the conduction and valence bands, giving rise to different effective
masses of electrons and holes, and thus different intercellular contributions to the valley moment. However, the
origins of the asymmetry have not been identified. As the result was sample dependent, it was suggested that natural
candidate for influencing the band structure is strain or doping. However, the work by  Li et al.\cite{Li2014}
convincingly demonstrates that doping has no influence on the valley splitting. This leaves only strain, as a possible
candidate to influence the valley Zeeman splitting, in line with theoretical predictions showing that strain leads to
asymmetry of the masses in the valence and the conduction band.\cite{Rostani15,Scalise14}

In this paper, we present polarization resolved magneto optical absorption measurements in the magnetic field up to
65~T on large area chemical vapor deposition (CVD) grown epitaxial monolayer molybdenum disulfide (MoS$_{2}$) and
molybdenum diselenide (MoSe$_{2}$) samples. Using $\sigma^\pm$ circularly polarized light we can individually address
absorption to the $\pm K$ valleys. In contrast to photoluminescence measurements, which generally probe only the
A-exciton, absorption provides easy access to the higher energy B-exciton which arises due to the large spin-orbit
splitting of the valence band in TMDs. At low temperatures ($T\simeq2$\,K) and in a magnetic field, both excitons
exhibit a large splitting of the $\sigma^\pm$ transitions with an effective valley g-factor $g_v \simeq -4$ in
agreement with previous magneto optical investigations of the A-exciton in exfoliated single layer
TMDs.\cite{Wang15,Mitioglu2015,Srivastava2015,Li2014} The similar values for the valley g-factor of the A and B
excitons is in line with band structure calculations.\cite{Ramasubramaniam12,Xu2014} We find that for both excitons the
value of the valley g-factor is approximately independent of the temperature. In CVD grown samples, strain is naturally
induced by the growth at high temperatures.\cite{Dumcenco2015,Liu14} Our results demonstrate that the strain induced by
the different coefficients of thermal expansion of the TMD and the sapphire substrate has a negligible influence on the
electronic properties.

\section{Experimental details}

The large area monolayer molybdenum disulfide (MoS$_2$) and molybdenum diselenide (MoSe$_2$) samples have been obtained
by the CVD method on highly polished sapphire substrates.~\cite{Dumcenco2015} Prior to the growth, the substrates were
cleaned by acetone/isopropanol/DI-water and further annealed at $1000^\circ$C in air for one hour. The growth process
is based on the gas-phase reaction between MoO$_3$ ($\geq 99.998\%$ purity, Alfa Aesar) and sulfur/selenium evaporated
from solid phase ($\geq 99.99\%$ purity, Sigma Aldrich). A crucible, containing $\sim 5$mg MoO$_3$ with the sapphire
substrates placed face-down above it, was loaded into a $32$\,mm outer diameter quartz tube placed in a three-zone
furnace. A second crucible located upstream from the growth substrates contained $350$\,mg of sulfur or 150 mg of
selenium. Ultrahigh-purity argon (Ar) was used as the carrier gas, and CVD growth was performed at atmospheric
pressure. The recipe for the MoS$_{2}$ growth is as follows; ramp the temperature to $300^{\circ}$C ($200$ sccm of Ar
flow) and set $300^{\circ}$C for 10 minutes, ramp to $700^{\circ}$C with $50^{\circ}$C min$^{-1}$ rate (10 sccm of Ar)
and set $700^{\circ}$C for 10 minutes, cool down to $570^{\circ}$C and open the furnace for rapid cooling (increase the
Ar flow to 200 sccm). The initially triangular shaped monolayers of MoS$_{2}$ merge into a large-area continuous film
with typical dimensions of a few mm over $\simeq 1$\,cm. For MoSe$_2$, in addition to 10 sccm of Ar, 3 sccm of H$_2$
was introduced during 10 minutes growth at $700^\circ$C. More details concerning the growth can be found in the
supplementary information section of reference\,[\onlinecite{Dumcenco2015}].

Polarized-resolved magneto-optical measurements have been performed at different temperatures using 70\,T long-duration
pulsed magnet ($\sim$500ms duration). A tungsten halogen lamp was used to provide a broad spectrum in the visible and
near-infrared range. The absorption was measured in the Faraday configuration in which the light propagation vector
$\mathit{k}$ is parallel to the magnetic field $B$. Typical size of the spot was of the order of 200 $\mu$m which is
much smaller compared to the dimensions of the monolayer TMD film. The circular polarization optics which allows to
selectively probe the transitions in one of the valleys was introduced \textit{in-situ}. To detect the opposite
circular polarization, the magnetic field direction was reversed. In our work, the $\sigma\pm$ polarization was
arbitrarily assigned to have a negative valley g-factor in agreement with the literature. All spectra were normalized
by the incident intensity to produce absolute transmission spectra.

\section{Magneto optical absorption spectroscopy}
\subsection{Valley g-factors}

\begin{figure}[b!]
\begin{center}
\includegraphics[width= 7.5cm]{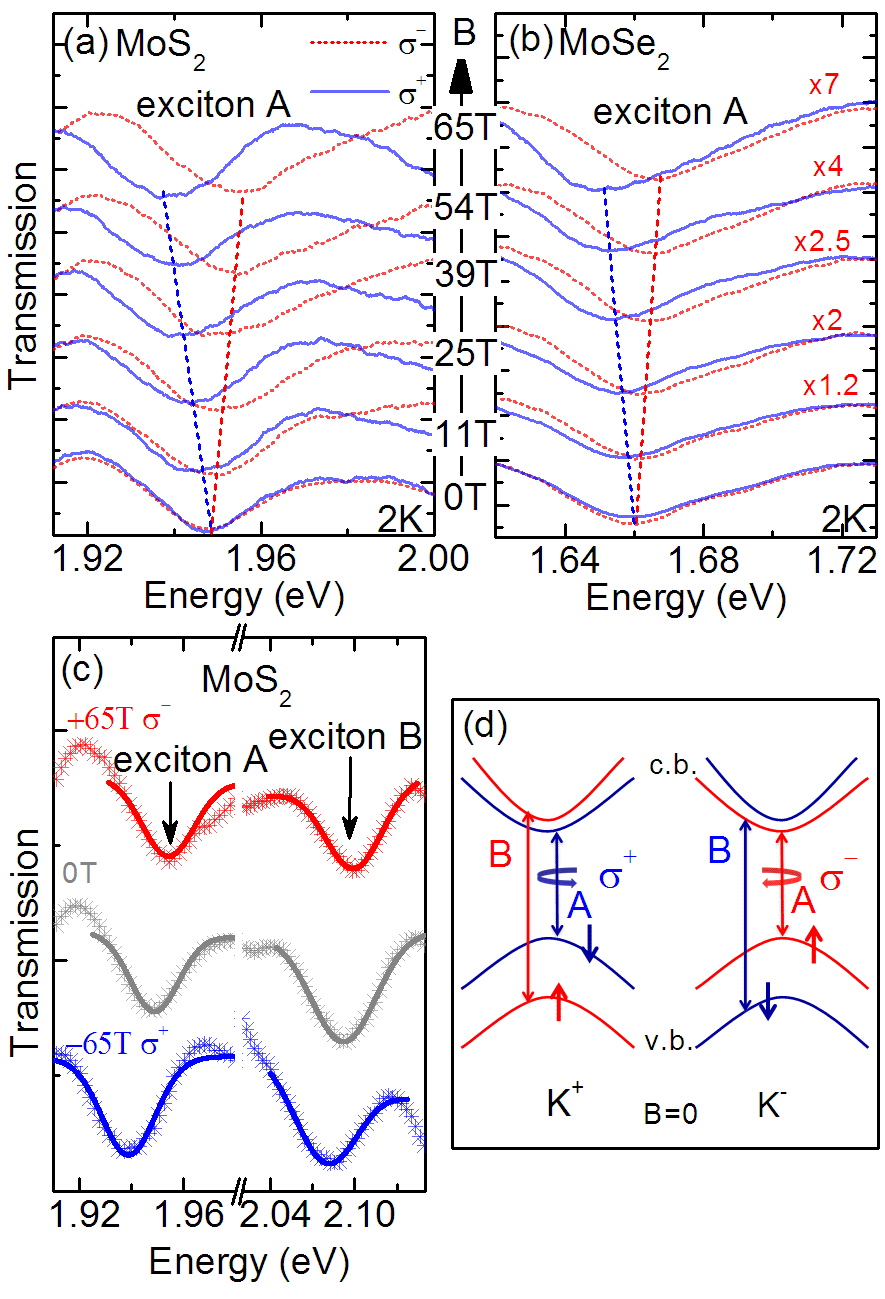}
\end{center}
\caption{(color online)(a-b) Typical low temperature transmission spectra for MoS$_{2}$ and MoSe$_{2}$ showing data
obtained for $\sigma^{+}$ and $\sigma^{-}$ polarization. (c) Example of the fitting Gaussian function to the spectra at
B=0 and $\pm$ 65T. (d) Schematic showing the optical selection rules and the shift of the bands in the magnetic
field.}\label{Fig1}
 \end{figure}

Representative low temperature magneto-transmission spectra obtained for a single layer MoS$_{2}$ and MoSe$_{2}$
showing A-exciton absorption are presented in Fig.~\ref{Fig1}(a-b) for $\sigma^{+}$ and $\sigma^{-}$ circular
polarization. For MoSe$_{2}$ each $\sigma^{-}$ spectra in magnetic field has been multiplied by a suitable numerical
factor to have a similar absorption intensity as the $\sigma^{+}$ spectra. The minima observed in all the spectra
occurs at an energy corresponding to the expected A-exciton absorption in both materials.\cite{Wang2012}

A clear splitting of both exciton transitions is observed which increases linearly with increasing the magnetic field
and reaches about 18~meV at the maximum applied field (65\,T). Such a splitting has been previously observed in PL
measurements at lower magnetic field in exfoliated
samples.~\cite{Li2014,MacNeill2015,Srivastava2015,Aivazian2015,Wang2012,Mitioglu2015} The valley splitting arises from
the opposite sign of the valley magnetic moment in the valence band. The relative magnetic field induced energy shift
of the valence and conduction band in each valley is schematically presented in Fig.~\ref{Fig1}(d). The dipole-allowed
transitions for the A and B-excitons are indicated by the vertical arrows. For both excitons $\sigma^{+}$ polarized
light couples to the $+K$ valley while $\sigma^{-}$ polarized light couples to the $-K$ valley. In the absence of a
magnetic field, the $\pm K$ transitions have identical energies for both the A and B-excitons. Applying a magnetic
field breaks the time-reversal symmetry, lifting the valley degeneracy, which splits the $\pm K$ ($\sigma^\pm$)
transitions. It is important to note that the schematic in the Fig.~\ref{Fig1}(d) is valid only for molybdenum
dichalcogenides. For the tungsten dichalcogenides, the order of the spin up/down conduction bands is
reversed.\cite{Kormanyos2015}.

\begin{figure}[t!]
 \begin{center}
 \includegraphics[width= 7.5cm]{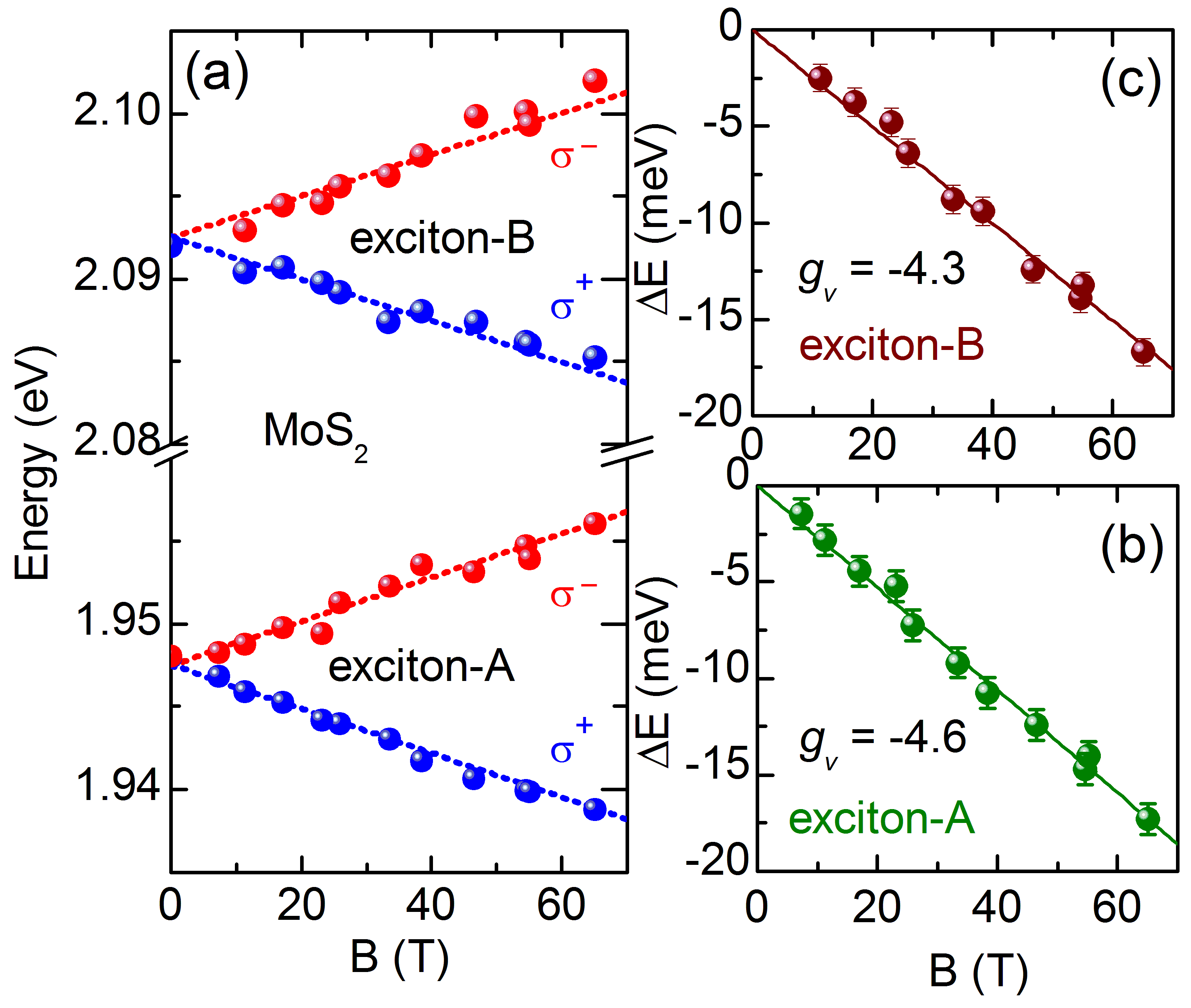}
 \end{center}
 \caption{(color online)
(a) Transition energies for the A and B-excitons in monolayer MoS$_{2}$ at $T=2$\,K. (b-c) The A and B-exciton valley
splitting at $T=2$\,K. The solid lines are linear fits used to extract the indicated valley g-factors. The broken lines
in (a) are the calculated evolution of the transition with magnetic field assuming a valley splitting of $\pm 0.5 g_v
\mu_B B$.}\label{Fig2}
 \end{figure}

\begin{table}[b!]
\caption{Summary of the temperature dependence of the valley $g$-factors for A and B-excitons in MoS$_{2}$ and the A
exciton in MoSe$_{2}$.}\label{Tab_t1}
\begin{center}
\begin{tabular}{c|c|c|c}
T(K) & MoS$_{2}$ A & MoS$_{2}$ B & MoSe$_{2}$ A\\
\hline
2 & $-4.6 \pm 0.1$& $-4.3 \pm 0.1$ & $-4.4 \pm 0.1$\\
77 & $-4.4 \pm 0.1$ & $-4.2 \pm 0.1$ & $-4.3 \pm 0.1$\\
120 & $-4.6 \pm 0.1$ & $-4.3 \pm 0.1$ & $-3.9 \pm 0.1$\\
\end{tabular}
\end{center}
\end{table}

To extract the exciton splitting energy at each magnetic field, the energy of the absorption line was determined by
fitting a Gaussian function. Examples of the fitted spectra at $B=0$ and $65$\,T for both circular polarizations are
shown in Fig.~\ref{Fig1}(c). The spectra are shown for MoS$_{2}$ in the energy range covering both the A and B
excitons. The energy of the A and B-excitonic transitions as a function of magnetic field in monolayer MoS$_{2}$ is
plotted in Fig.~\ref{Fig2}(a) for $\sigma^\pm$ polarizations. For both excitons, the energy of the transitions evolve
linearly with magnetic field.

The difference between the transition energy with $\sigma^{+}$ and $\sigma^{-}$ circular polarized light ($\Delta E =
E_{\sigma^+} - E_{\sigma^-}$) in magnetic field for both excitons is presented in Fig.~\ref{Fig2}(b-c). The exciton
valley splitting scales linearly with the magnetic field and is almost identical for both excitons. A linear fit to the
data gives $g_v \simeq -4.6 \pm 0.1$ and  $g_v \simeq -4.3 \pm 0.1$ for the A and B-excitons respectively. Similar
values for the exciton A were reported for exfoliated monolayer MoSe$_{2}$ samples using photoluminescence measurements
in low magnetic fields.~\cite{Li2014,MacNeill2015} The expected evolution of the transition energies in magnetic field,
calculated using the valley Zeeman splitting $\pm 0.5 g_v \mu_B B$ is indicated in Fig.\,\ref{Fig2}(a) by the broken
lines. The excellent agreement with the data confirms that within experimental error the splitting is symmetric with no
evidence for a diamagnetic shift or cyclotron-like free carrier contribution to the magnetic field evolution of the
transitions.

\begin{figure}[t!]
\begin{center}
\includegraphics[width= 7.5 cm]{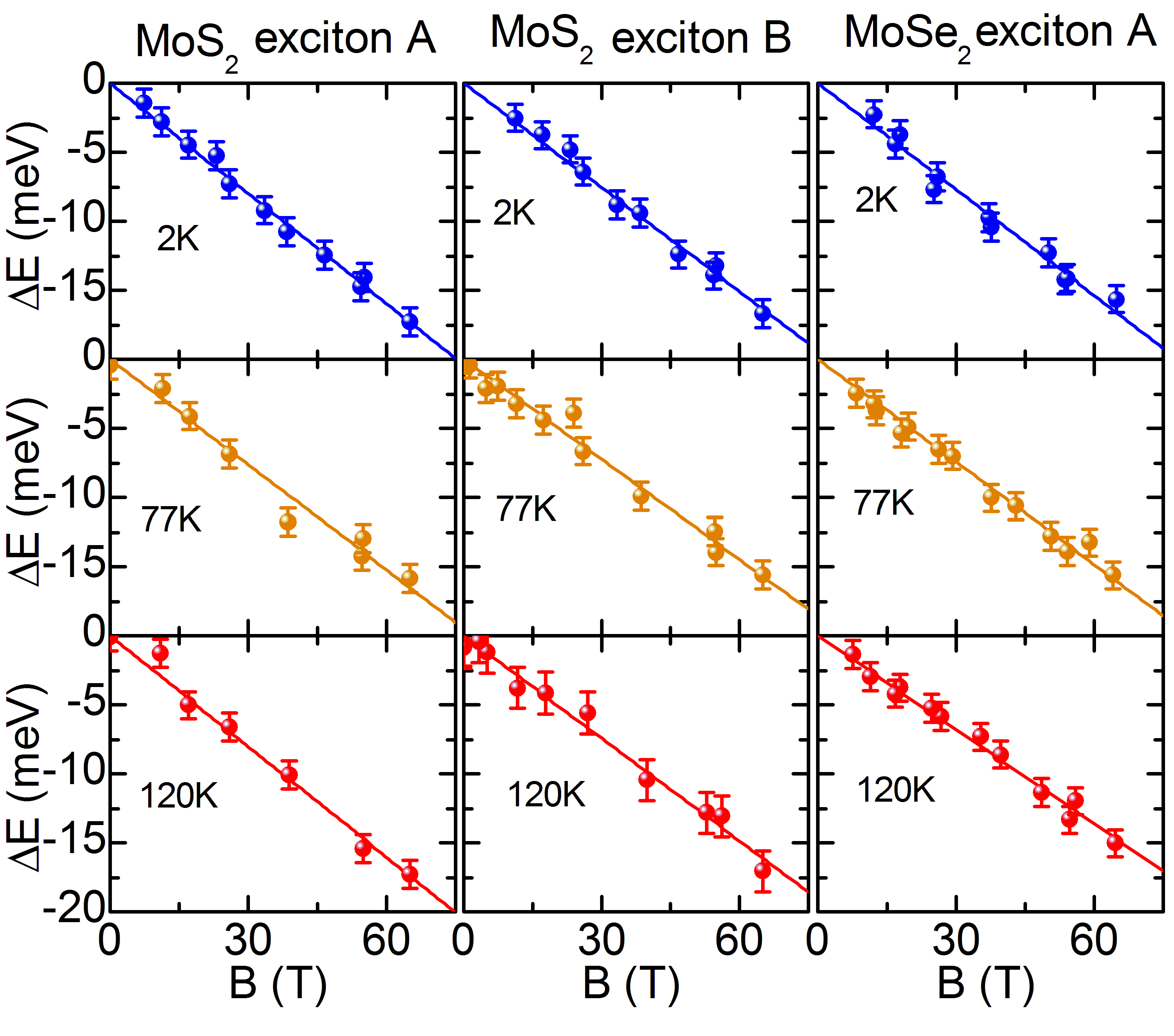}
\end{center}
\caption{(color online) The exciton valley splitting in monolayer MoS$_{2}$ and MoSe$_{2}$ for three different
temperatures. The solid lines are linear fits to the data made to extract the valley g-factors summarized in
Table\,\ref{Tab_t1}.}\label{Fig4}\end{figure}

 We have measured transmission spectra at $B=0$ and $B=\pm65$\,T
($\sigma^\pm$) for three different temperatures for the A and B-excitons in MoS$_{2}$ and A-exciton in MoSe$_{2}$
(exciton B is not resolved in our CVD MoSe$_{2}$ samples). The precise position of the exciton transitions as a
function of magnetic field was obtained by fitting Gaussian functions. The obtained splitting $\Delta E$ between the
$\sigma^{+}$ and $\sigma^{-}$ transitions is plotted in Fig.~\ref{Fig4} versus the magnetic field at three different
temperatures. The valley g-factors were extracted by linear fits to the data (solid lines). The values of the
temperature dependent valley g-factor are summarized in Table\,\ref{Tab_t1}. In MoS$_2$, for both excitonic transitions
the valley g factor is independent of the temperature within experimental error. For MoSe$_2$, where only the exciton A
is observed, the g-factor is constant within experimental error for $T\leq77$\,K and decreases by around 10\% at
$T=120$\,K.

\subsection{The influence of strain}

Strain modifies the ratio of the effective masses in the valence and the conduction bands giving rise to an
intercellular contribution to the valley splitting which then takes the form $\Delta E = 4 \mu_B B - 2 \Delta \alpha
\mu_B B$, where $\Delta \alpha = (1/m_{c} - 1/m_{v})$ and $m_{c},m_{v}$ are the effective masses in the conduction and
valence band in the units of the free electron mass. \cite{liu13,Aivazian2015} In principle $\Delta \alpha$ can be
calculated taking into account higher order corrections to the tight binding model. Estimations vary between $0.2$ to
$1.1$ depending if only nearest neighbor (NN) or next nearest neighbor (NNN) hopping parameters are taken into account
in three band tight binding model~\cite{liu13,Aivazian2015}. CVD grown samples, are naturally
strained\cite{Dumcenco2015,Liu14}, which results in the modification of the band structure, in particular the ratio of
the effective masses in the valence and conduction bands\cite{Conley13,Yue12,Scalise14,Rostani15}.

\begin{figure}[hb!]
\begin{center}
\includegraphics[width= 6.5 cm]{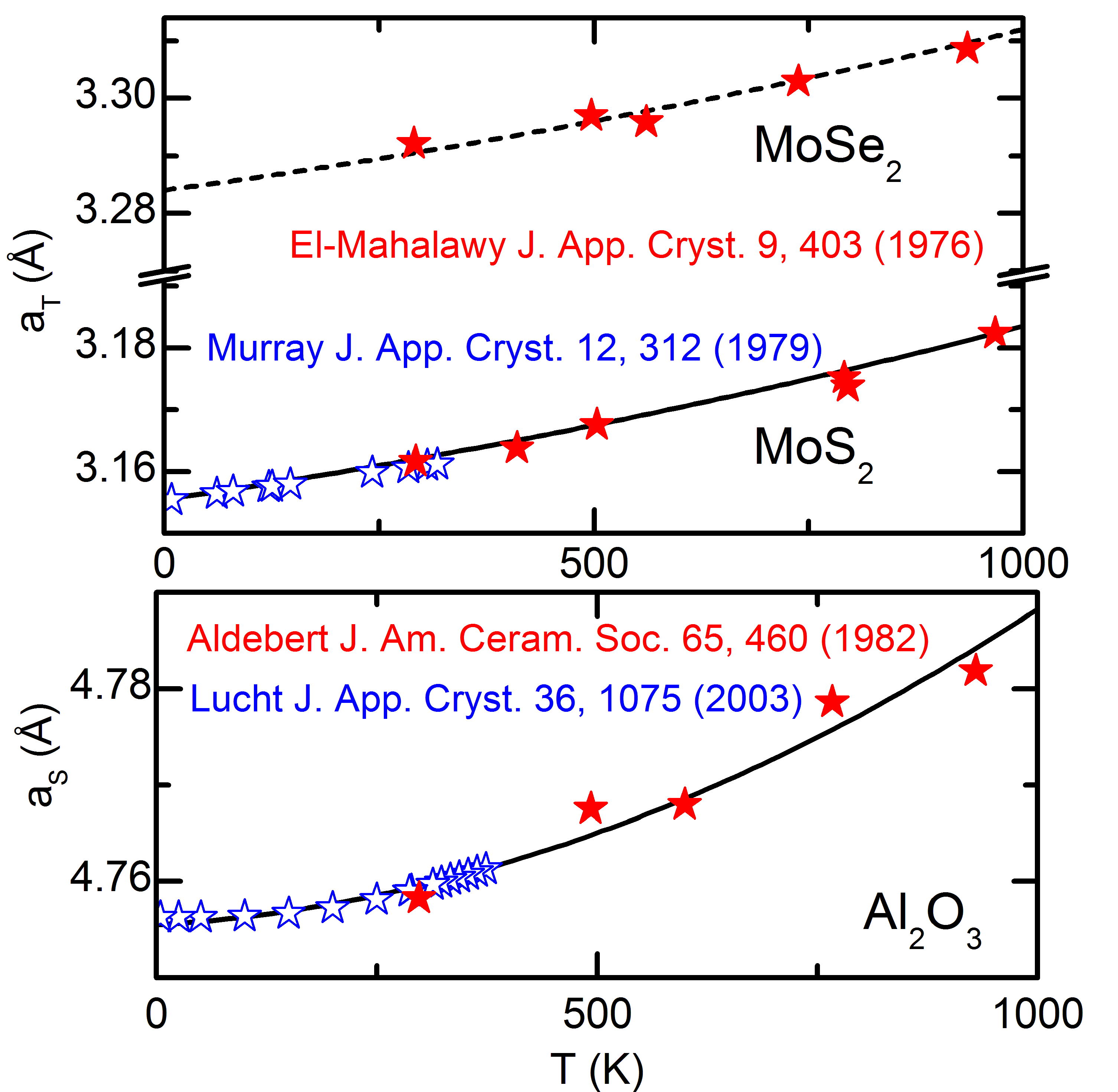}
\end{center}
\caption{(color online) Temperature dependence of the measured $a$ lattice constant of bulk MoS$_{2}$, MoS$e_{2}$ and
sapphire (Al$_2$O$_3$) taken from the literature.\cite{Alderbert1982,El-Mahalawy82,Lucht2003,Murray79} Note that for
the low temperature MoS$_2$ data, scatter was reduced by taking points corresponding to the linear fit to the data in
reference [\onlinecite{Murray79}]. The samples were grown at $T=973$\,K. The solid lines are fits to the data using
second order polynomials. The dashed line is the MoS$_{2}$ fit shifted vertically to coincide with the MoS$e_{2}$ high
temperature data.}\label{LatticeConstant}
\end{figure}

To estimate the strain we need to know the temperature dependence of the lattice constant $a$ of the bulk TMDs and
sapphire from the lowest measurement temperature up to the growth temperature ($700^{\circ}$C$ \equiv 973$\,K) at which
the TMD monolayer is assumed to be unstrained. In Fig.\,\ref{LatticeConstant} we plot the $a$ lattice constant
assembling published data in the literature\cite{Alderbert1982,El-Mahalawy82,Lucht2003,Murray79} in order to span the
temperature range of interest for MoS$_2$, MoSe$_2$ and sapphire. The solid lines are second order polynomial fits
which we use to calculate the strain versus temperature. For MoSe$_2$ we were unable to find any published data below
room temperature. Fortunately, the high temperature data suggests that the temperature dependence of MoS$_2$ and
MoSe$_2$ are almost identical. We therefore use the fitted temperature dependence of MoS$_2$ which has been shifted
vertically (broken line).

The strain is by definition,
\begin{equation}
\varepsilon(T) = \frac{a^{'}_{T}(T) - a_{T}(T)}{a_{T}(T)},\nonumber
\end{equation}
where $a^{'}_{T}(T)$ is the lattice constant of the TMD grown on sapphire and $a_{T}(T)$ is the lattice constant of the
unstrained bulk TMD. Assuming that the TMD monolayer is constrained to follow the thermal contraction of the sapphire
substrate when the sample is cooled from the growth temperature we can write,
\begin{equation}
\frac{a^{'}_{T}(T)}{a_{T}(973)} = \frac{a_{S}(T)}{a_{S}(973)},\nonumber
\end{equation}
where $a_{S}(T)$ is the lattice constant of the sapphire substrate. Thus, the strain is given by,
\begin{equation}
\varepsilon(T) = \frac{a_{S}(T) a_{T}(973)}{a_{S}(973) a_{T}(T)} - 1,\nonumber
\end{equation}
which can be calculated using the polynomial approximations for the evolution of the lattice constants with
temperature.

\begin{figure}[ht!]
\begin{center}
\includegraphics[width= 6.5 cm]{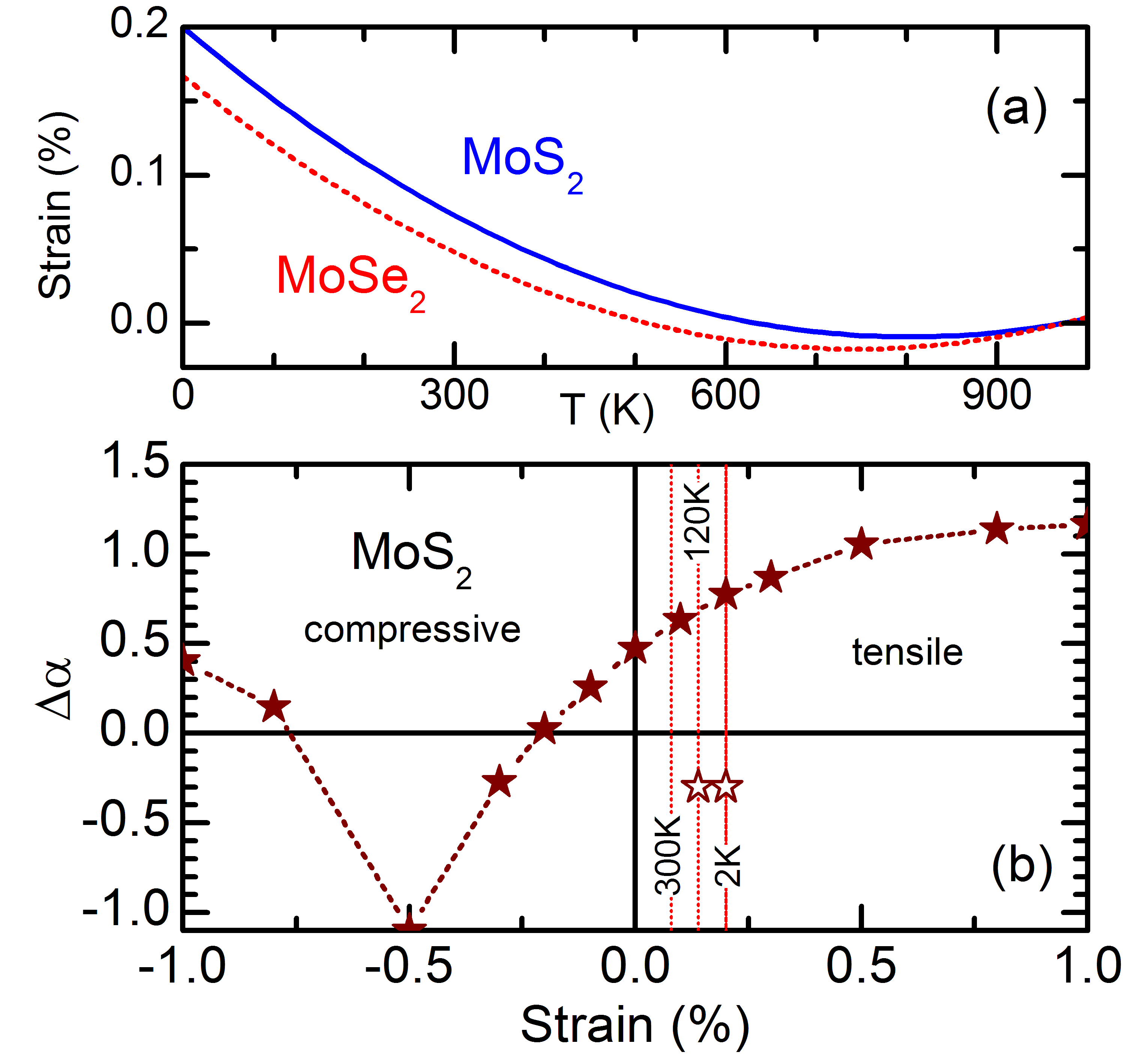}
\end{center}
\caption{(color online) (a) The calculated strain as a function of temperature due to the different thermal contraction
of the TMD monolayer and the Al$_2$O$_3$ substrate. (b) Closed symbols show the calculated intercellular correction
coefficient $\Delta \alpha$ for MoS$_{2}$ as a function of the strain. The open symbols show the required values of
$\Delta \alpha$ to agree with the measured valley g-factors.}\label{Strain}
\end{figure}

The calculated strain for MoS$_2$ and MoSe$_2$ is plotted in Fig.\,\ref{Strain}(a). For both TMDs the strain remains
negligibly small as the sample is cooled from the growth temperature to $600$\,K. Below this temperature the tensile
strain progressively increases reaching 0.2\% in MoS$_2$ at low temperature. For MoSe$_2$ the strain is slightly
smaller reaching a maximum value of $\simeq 0.17$\% at $T=0$\,K.

In order to estimate the intercellular correction to the valley magnetic moment we have calculated $\Delta \alpha$
using the electron and hole effective masses at the $K$-point in MoS$_2$, calculated as a function of strain, using
density-functional theory (DFT).\cite{Scalise14} The result is plotted as closed symbols in Fig.\,\ref{Strain}(b). The
vertical broken lines indicate the expected strain in MoS$_2$ at $2$K, $120$K and room temperature. While $\Delta
\alpha$ varies little over the range of strain of interest, in line with the observation that the valley g-factor
remains almost unchanged with temperature, the value of $\Delta \alpha$, notably its sign, is not in agreement with the
measured g-factors with $g_v < -4$ which implies that $\Delta \alpha$ should be negative. The open symbols indicated
the required values of $\Delta \alpha$ to have agreement with the experimental valley g-factors. The failure of the DFT
calculations to correctly predict the intercellular correction is not unexpected as excitons in TMDs are highly
localized in real space, and thus delocalized in $k$-space. This can in principle be taken into account by averaging
the electron and hole effective masses over momentum space in the vicinity of the $K$-points, which leads to a negative
value of $\Delta \alpha$ in agreement with experiment.\cite{Srivastava2015} However, such calculations are beyond the
scope of this work.

\begin{figure}[t!]
\begin{center}
\includegraphics[width= 6.5 cm]{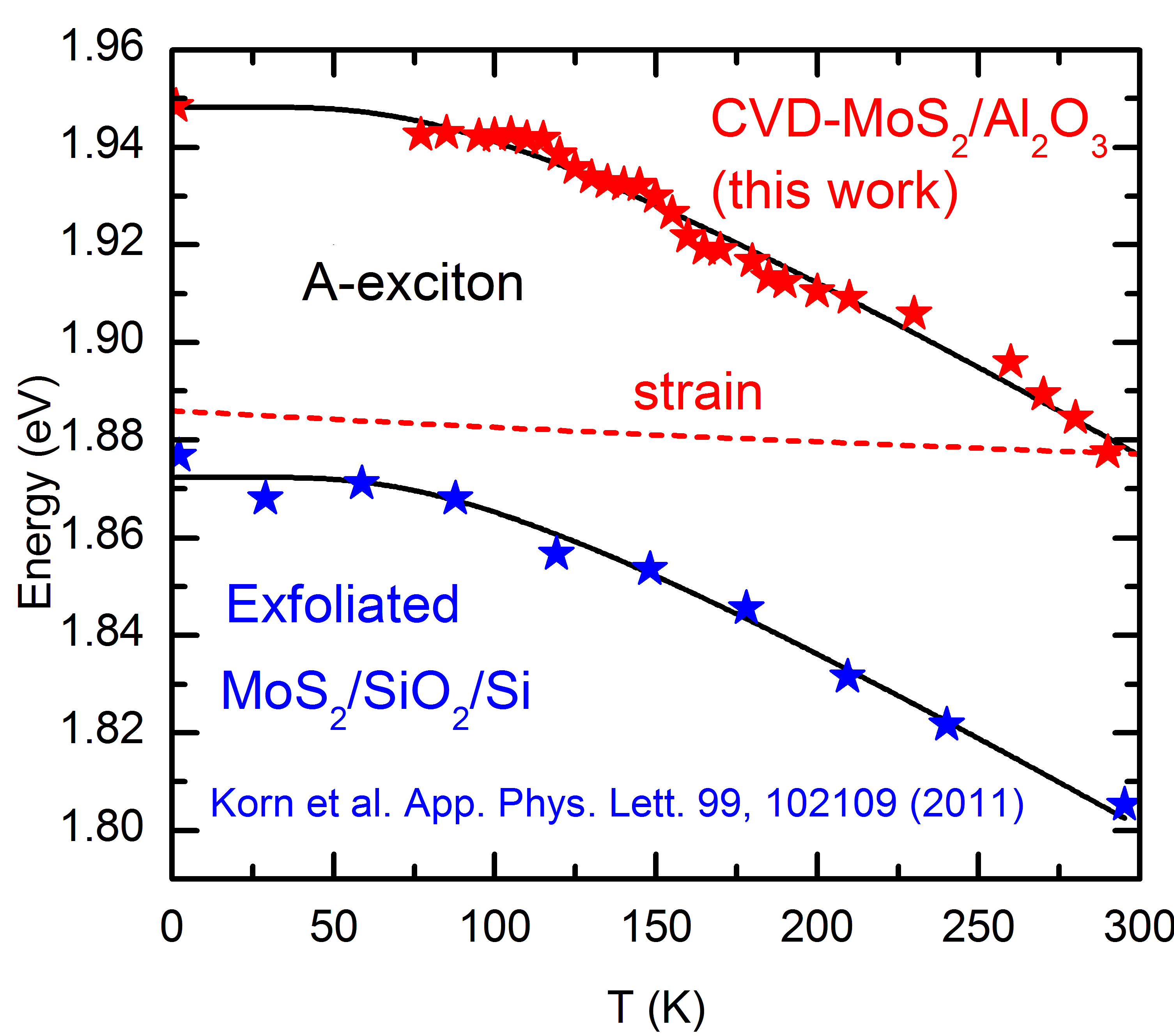}
\end{center}
\caption{(color online) The measured evolution of the A-exciton absorption in CVD MoS$_{2}$ on Al$_2$O$_3$. Data from
the literature\cite{Korn11} for the A-exciton emission from exfoliated MoS$_{2}$ on SiO2/Si substrate is shown for
comparison. The solid lines are the fitted temperature dependence as described in the text. The broken line shows the
calculated small contribution of the changing strain to the temperature dependence of CVD MoS$_{2}$ on
Al$_2$O$_3$.}\label{TdepGap}
\end{figure}

Finally, the negligible influence of strain in CVD grown TMDs, is confirmed by the temperature dependence of the
A-exciton absorption in MoS$_2$ plotted in Fig.\,\ref{TdepGap}. We use the single oscillator model of O'Donnell and
Chen to model the temperature dependence of the band gap.\cite{Donnell91} The solid line is a fit to the data using
\begin{equation}
E(T) = E(0) - S \braket{\hbar \omega} \left[\coth\left(\frac{\braket{\hbar \omega}}{2KT}\right) - 1\right],\nonumber
\end{equation}
where $\braket{\hbar \omega}=24.25$\,meV is the average phonon energy, $S=2.29$ is a dimensional coupling constant and
$E(0)=1.948$\,eV is the low temperature band gap minus the exciton binding energy. The fit is excellent suggesting that
strain plays little role in the observed temperature dependence. The expected variation of the band gap due to the
change in strain with temperature, calculated from the measured $\simeq70$meV/\%\,strain red shift in monolayer
MoS$_2$,\cite{He2013} is shown by the dashed line in Fig.\,\ref{TdepGap}. Clearly the expected strain-induced
$\simeq8$\,meV change in the band gap is small compared to the observed $\simeq 70$\,meV change with temperature.

For comparison, we plot the energy of the A-exciton emission in exfoliated MoS$_2$ on a SiO$_2$/Si substrate taken from
the literature.\cite{Korn11} The energy of the emission is systematically shifted by $\simeq 72$\,meV due to the
different dielectric environment (exciton binding energy). The solid line through the data is calculated using the same
parameters as for the CVD MoS$_2$ except for $E_0=1.872$\,eV which is shifted due to the increased exciton binding
energy. The excellent agreement with experiment demonstrates that exfoliated and CVD MoS$_2$ have the same temperature
dependence of the band gap, further confirming the negligible role played by strain in the latter.

\section{Conclusion}

We have investigated large-area monolayer MoS$_{2}$ and MoSe$_{2}$ samples, grown by CVD on sapphire, in high magnetic
fields using optical absorption spectroscopy.  The exciton valley splitting scales linearly with the magnetic field. In
MoS$_{2}$  the extracted low temperature ($2$\,K) valley g-factors are $g_{v}\simeq -4.5 \pm 0.1$ for the A-exciton and
$g_{v}\simeq -4.3 \pm 0.1$ for the B-exciton. In MoSe$_{2}$ for which only the A-exciton was observed we find
$g_v=-4.4\pm0.1$ at low temperatures. In both TMDs the g-factor is almost independent of temperature over the available
measurement range (2-120\,K). The strain present at low temperature $\simeq 0.2$\% in our CVD grown TMDs has little
effect on the electronic properties. The low temperature valley g-factors and the temperature dependence of the gap are
identical to unstrained exfoliated MoS$_2$ on SiO$_2$/Si substrates. This suggests that the $\simeq0.2$\% tensile
strain, which is naturally present in large area CVD grown Mo based TMDs on sapphire, does not represent any serious
impediment for device applications.

During preparation of the manuscript, we became aware of similar work on CVD grown WS$_{2}$ and MoS$_{2}$ monolayers by
the NHMFL-Los Alamos group.~\cite{Stier2015}

\begin{acknowledgments}
This work was partially supported by ANR JCJC project milliPICS, the Region Midi-Pyr\'en\'ees under contract MESR
13053031, STCU project 5809 and the Swiss SNF Sinergia Grant no. 147607.
    \end{acknowledgments}


%

\end{document}